\documentclass[]{aa}
\usepackage{txfonts}
\usepackage{natbib}
\usepackage{amssymb}
\bibpunct{(}{)}{;}{a}{}{,}
\usepackage{epsf}
\usepackage{graphicx}
\usepackage{epsfig}
\usepackage{graphics}
\usepackage{subfigure}
\usepackage[english]{babel}
\usepackage{rotating}
\usepackage{color}

\newcommand{\lL}{\ifmmode \log \frac{L}{L_{\sun}} \else $\log \frac{L}{L_{\sun}}$\fi}

\newcommand{\msun}{M$_{\sun}$}

\begin{document}

\title{Quantitative spectral classification of Galactic O stars}
\author{F. Martins\inst{1}
}
\institute{LUPM, Universit\'e de Montpellier, CNRS, Place Eug\`ene Bataillon, F-34095 Montpellier, France  \\
}

\offprints{Fabrice Martins\\ \email{fabrice.martins@umontpellier.fr}}

\date{Received / Accepted }

\abstract{Our goal is to provide a quantification of several spectral classification criteria for O stars.
We collect high-spectral resolution spectra of 105 Galactic O-type stars from various archives. We measured equivalent widths of classification lines. We defined average values of classification criteria for given spectral types and luminosity classes.
We find that the ratio \ion{He}{i}~4471 to \ion{He}{ii}~4542 well matches the published ratios for spectral types. We have quantified equivalent width ratios of helium and silicon lines among O8-O9.7 stars to refine spectral class typing in this spectral range. We present quantitative criteria to separate between luminosity class V, IV-III-II (grouped), and I among O3-O8.5 stars, mainly based on the strength of \ion{He}{ii}~4686. We find that these criteria also define very well the f, (f), and ((f)) classes for O3-O7.5 stars. Among O9-O9.7 stars we quantify the ratios of \ion{He}{ii}~4686 to \ion{He}{i}~4713 and \ion{Si}{iv}~4089 to \ion{He}{i}~4026 for all luminosity classes.
The tabulated values of the classification criteria should help classify any new O-type stars. The final step of the classification process should rely on a direct comparison to standard stars of the assigned spectral type or luminosity class.}


\authorrunning{F. Martins}
\titlerunning{Quantitative spectral classification of O stars}

\maketitle

\section{Introduction}
\label{s_intro}

Massive stars have initial masses higher than about 10 \msun. They are born as O and B stars and evolve into blue, yellow and red supergiants, Luminous Blue Variables or Wolf-Rayet stars. They end their lives as core-collapse supernovae of type II, Ib or Ic, sometimes associated with a long-soft gamma-ray burst. On and just after the main sequence they appear as O or early B stars. O stars are characterized by the presence of \ion{He}{ii} lines in their optical spectra. The MK classification system \citep{mkk43,mk73} introduced criteria to separate O stars with different ratios of \ion{He}{i} to \ion{He}{ii} lines. Earlier (later) spectral types have stronger \ion{He}{ii} (\ion{He}{i}) lines. The second dimension of the MK classification -- luminosity class -- was based on the relative strength if \ion{Si}{iv} to \ion{He}{i} lines.

The classification scheme was refined by \citet{walborn71a,walborn71b,walborn72,walborn02}. New spectral classes were introduced based on both the improved sensitivity of detectors which allowed the detection of weak lines and the addition of nitrogen and carbon lines as classification diagnostics. In particular spectral types O2 to O3.5 where introduced among the earliest O stars based on nitrogen and weak \ion{He}{i} lines. The classification scheme developed in the above series of papers relies on the definition of standards for each spectral type and luminosity class: any newly discovered O star is classified according to its spectral similarities to these standards. A catalog of reference optical spectra for the standards stars was presented by \citet{wf90}.

In parallel, \citet{ca71,cl74,cf77} and subsequently \citet{mathys88,mathys89} developed a classification based on measurement of equivalent widths (EWs). They defined ranges for such EWs or ratios of EWs within which a given spectral type or luminosity class was assigned. For spectral types a quantitative criterion was defined for the full range, from O3 to O9.7. However only for the latest spectral types could they quantitatively define luminosity classes. More recently, \citet{arias16} defined a quantitative criterion to distinguish normal dwarfs from Vz stars (stars with very strong \ion{He}{ii} absorption lines). 

These approaches (standard stars, quantitative criteria) are complementary. However since the advent of digital optical spectra, the scheme based on standard stars has been updated the most. The Galactic O Stars Spectroscopic Survey \citep{ma11} collected medium-resolution optical spectra for essentially all the Galactic O stars. Based on this unique database, \citet{sota11} and \citet{sota14} presented refined classification criteria for both spectral types and luminosity classes. In particular, they introduced the new spectral class O9.2. But these criteria remain qualitative. For instance, luminosity classes among early O stars rely on whether \ion{He}{ii}~4686 is in emission (supergiants), in strong absorption (giants), or in very strong absorption (dwarfs) -- see Table~5 in \citet{sota11}. A direct comparison to standard stars is thus mandatory to assess the strength of \ion{He}{ii}~4686. This is usually not straightforward since broadening mechanisms, especially rotational broadening, affect the strength of any absorption line.

To partially reduce these difficulties, we present a quantitative study of classification criteria for spectral classes and luminosity classes among O stars. We used high-resolution optical spectra to measure equivalent widths (and ratios of EWs) for selected classification diagnostics. The results should facilitate the process of spectral classification of newly discovered stars (or of theoretical spectra). Once this is done a final comparison to standard stars spectra, for which physical parameters have been determined by \citet{hol17}, should validate the classification process.

\section{Method}
\label{s_method}

Our method relies on the measurement of equivalent widths for key classification lines. For that, we have used high spectral resolution spectra from various archives and sources:

\begin{itemize}

\item The \emph{SOPHIE} archive\footnote{\url{http://atlas.obs-hp.fr/sophie/}} \citep{SOarchive}. \emph{SOPHIE} is an \'echelle spectrograph mounted on the 193cm telescope of the Haute Provence Observatory \citep{bouchy13}. It covers the 3872-6943\AA\ wavelength range with a spectral resolution of 40000. All the \emph{SOPHIE} data used in the present paper have been obtained between 2014 and 2015 and have been described by \citet{herve15}. Archive data result from reduction with the \emph{SOPHIE} pipeline \citep{bouchy09} adapted from the ESO/HARPS software.

\item The \emph{ELODIE} archive\footnote{\url{http://atlas.obs-hp.fr/elodie/}} \citep{ELarchive}. \emph{ELODIE} was the precursor of \emph{SOPHIE} on the 193cm telescope of the Haute Provence Observatory. It covered the 3850-6800\AA\ wavelength range at a resolution of 42000. Details of the automated data reduction can be found on the instrument web page (see above).

\item The CFHT Science Archive at the Canadian Astronomical Data Center\footnote{\url{http://www.cadc-ccda.hia-iha.nrc-cnrc.gc.ca/en/cfht/}}. Data from the spectro-polarimeter \emph{ESPaDOnS} have been used. They cover the 3700-10500\AA\ wavelength range with a spectral resolution of 68000. Data reduction was performed automatically with the Libre Esprit software \citep{donati97}. See \citet{wade16} for more information on the data. 

\item The PolarBase archive of \emph{NARVAL} spectra\footnote{\url{http://tblegacy.bagn.obs-mip.fr/narval.html}} \citep{petit14}. \emph{NARVAL} is a twin of \emph{ESPaDOnS} mounted on the Telescope Bernard Lyot of the Pic du Midi Observatory. Data reduction was performed similarly to \emph{ESPaDOnS} data. 

\item The ESO archive\footnote{\url{http://archive.eso.org/cms.html}} of \emph{FEROS} spectra collected by the OWN survey \citep{barba10}. \emph{FEROS} is operated on the La Silla 2.2m telescope. It has a resolving power of 48000 and covers the 3500-9200\AA\ wavelength range. Data were automatically reduced using the ESO \emph{FEROS} pipeline. We also retrieved \emph{HARPSPol} spectra collected on the La Silla 3.6m telescope. They cover the range 3780-6910\AA\ with a resolution of 115000. 
  
\end{itemize}

\noindent The list of stars and instrumental resources is given in Tab.\ \ref{tab_obs}. A total of 105 spectra has been collected. In practice we have measured the EWs of the following lines: \ion{He}{i}~4026, \ion{Si}{iv}~4089, \ion{He}{i}~4144, \ion{He}{i}~4388, \ion{He}{i}~4471, \ion{He}{ii}~4200,  \ion{He}{ii}~4542, \ion{Si}{iii}~4552, \ion{N}{iii}~4630-34-40, \ion{He}{ii}~4686, and \ion{He}{i}~4713.

\section{Results}
\label{s_res}

In this section we present the quantification of classification criteria for spectral types and luminosity classes. 

\subsection{Spectral types}
\label{s_st}

\begin{figure}[]
\centering
\includegraphics[width=9cm]{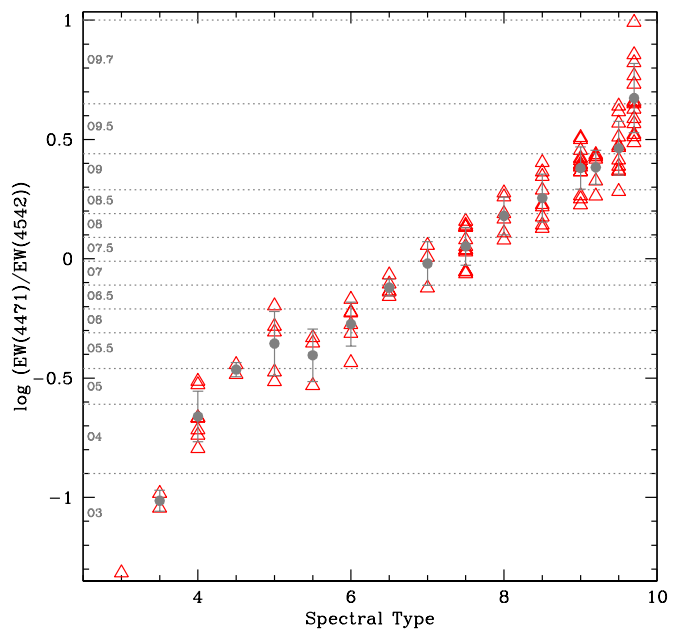}
\caption{Logarithm of the ratio of the equivalent width of \ion{He}{i}~4471 to the equivalent width of \ion{He}{ii}~4542 as a function of spectral type for our sample (red triangles). Gray points and error bars indicate the average and dispersion. Gray dotted lines show the limits for spectral types O3 to O9.7 defined by \citet{ca71} and \citet{mathys88,mathys89}.}
\label{ew_st_obs}
\end{figure}

Figure\ \ref{ew_st_obs} shows the classical spectral type criterion: the ratio of EW(\ion{He}{i}~4471) to EW(\ion{He}{ii}~4542) \citep{ca71,mk73,mathys89}. We note that the O9.2, O4.5 and O3.5 spectral types are not defined in the Conti \& Alschuler / Mathys classification scheme. Our average equivalent widths ratios fall nicely in the range defining each spectral type. The only exception is spectral O5 for which there is a large dispersion and an average ratio similar to the O5.5 range. The reason for this mismatch is unclear. ALS~2063 is classified as Ifp indicating peculiar lines and HD~93843 is variable: these peculiarities partly explain the large scatter in EW ratios. Removing both stars from the sample decreases the EW ratio down to the limit between spectral types O5 and O5.5. Other than that, O3.5 stars fall in the O3 range of Conti \& Alschuler / Mathys and O4.5 stars have EW ratios between O4 and O5-O5.5.
O9.2 stars have on average the same ratio of EW(\ion{He}{i}~4471) to EW(\ion{He}{ii}~4542) as O9 stars.

\begin{figure*}[]
\centering
\includegraphics[width=0.32\textwidth]{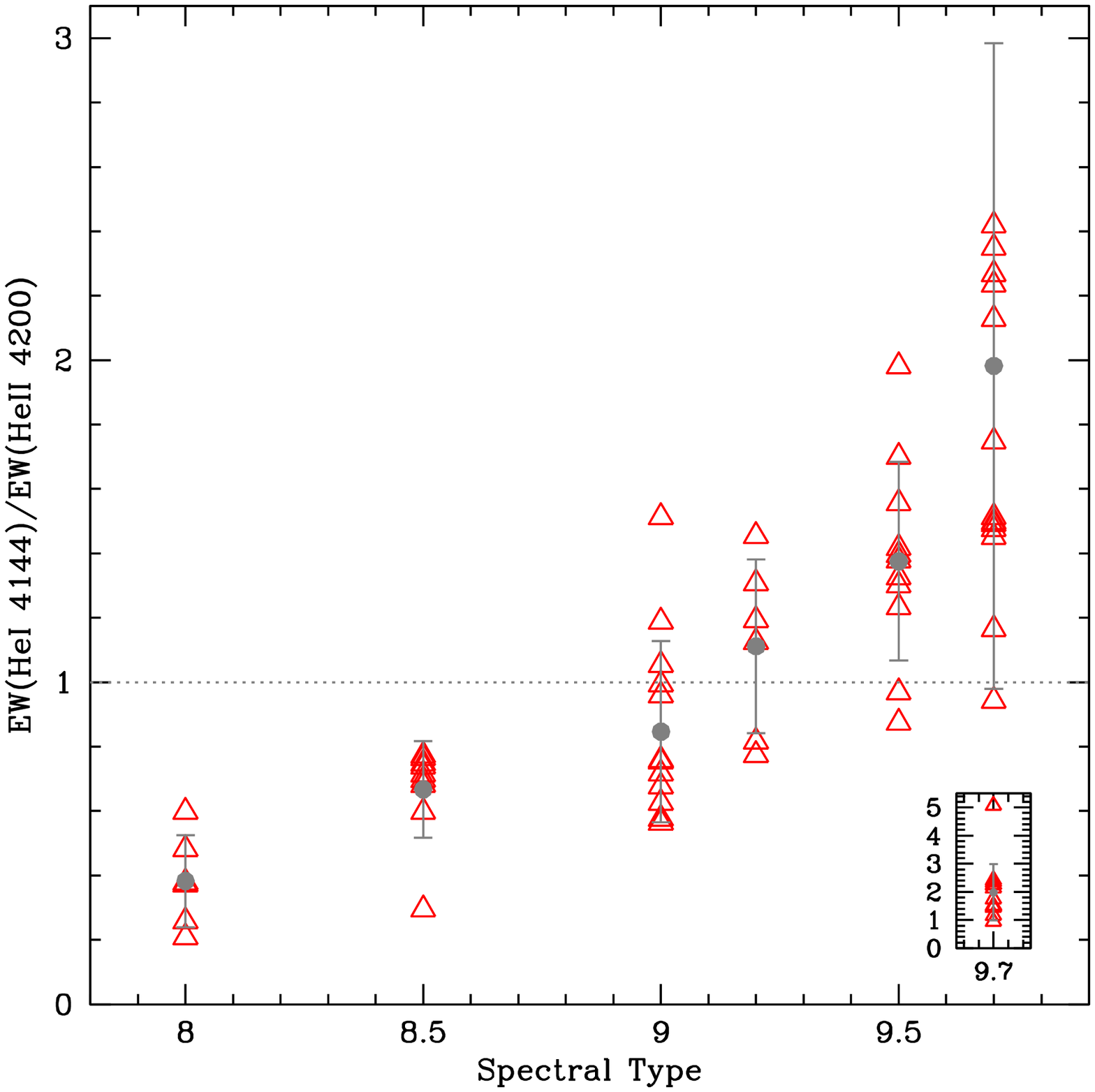}
\includegraphics[width=0.32\textwidth]{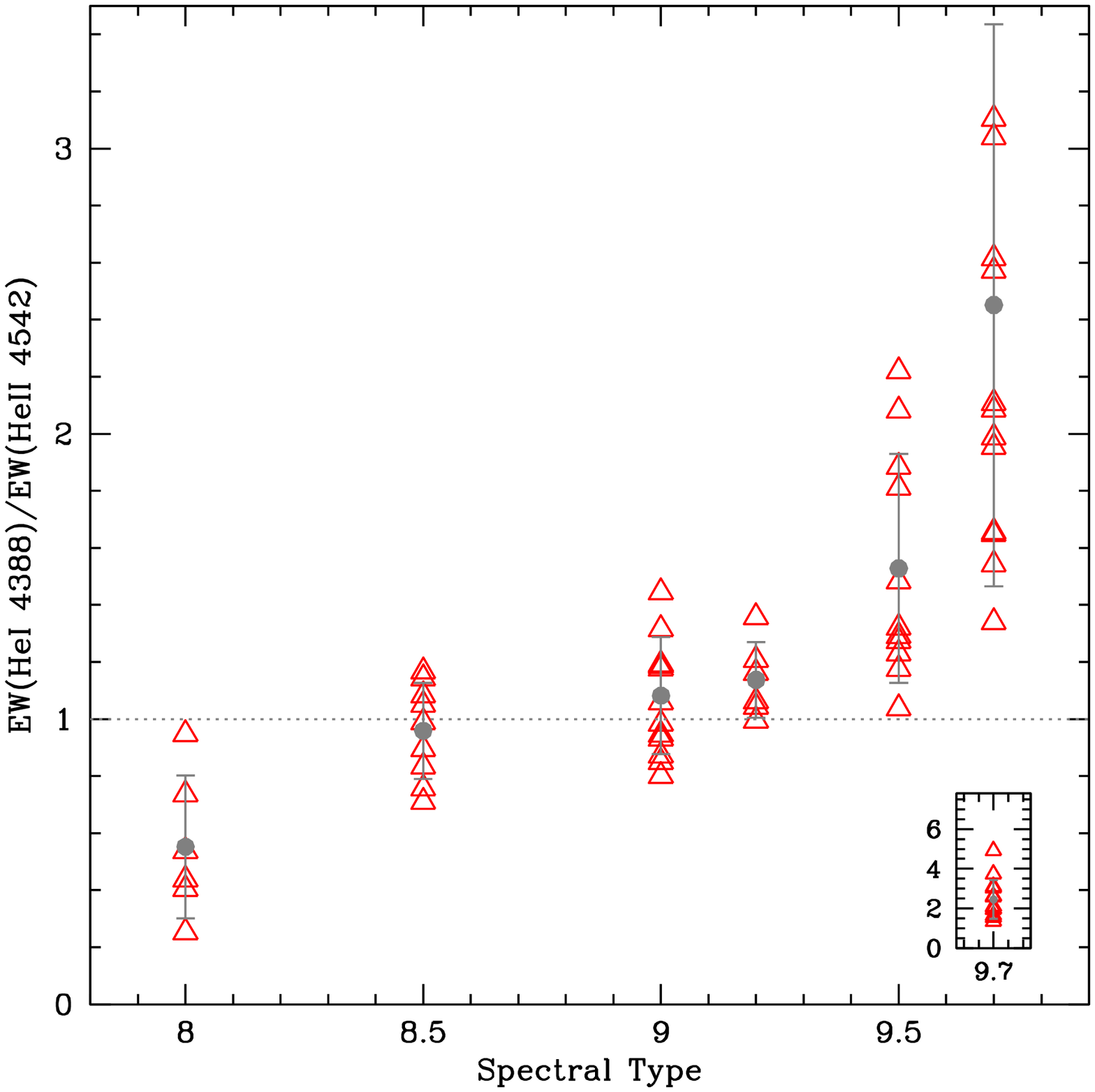}
\includegraphics[width=0.32\textwidth]{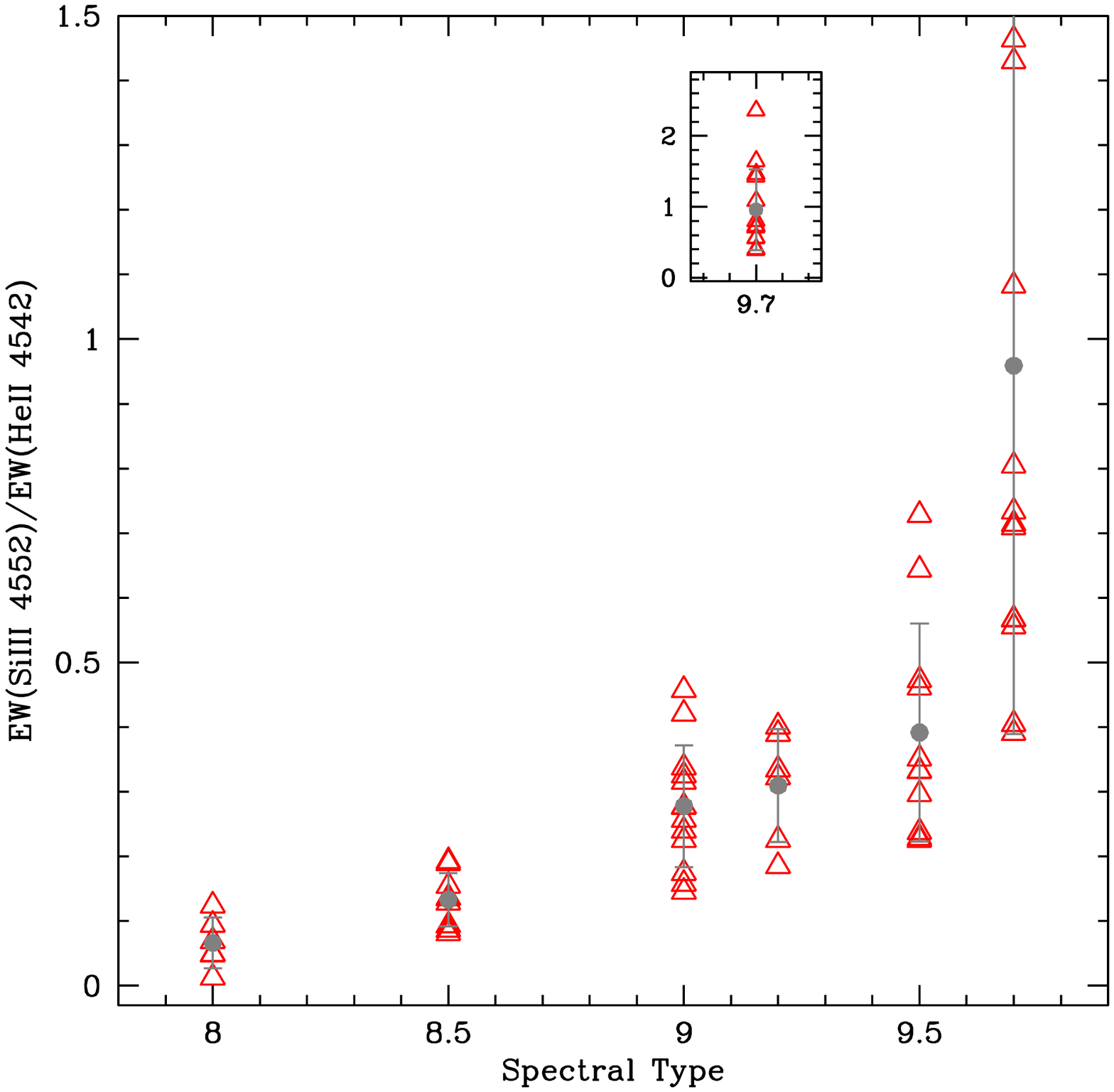}
\caption{Ratios EW(\ion{He}{i}~4144)/EW(\ion{He}{ii}~4200) (left panel), EW(\ion{He}{i}~4388)/EW(\ion{He}{ii}~4542) (middle panel panel), and EW(\ion{Si}{iii}~4552)/EW(\ion{He}{ii}~4542) (right panel) as a function of spectral type for O8-O9.7 stars (red triangles). Gray points and error bars indicate the average and dispersion. The gray dotted lines of the left and middle panels indicate a ratio of 1.0.}
\label{ew_st89_obs}
\end{figure*}

In Fig.\ \ref{ew_st89_obs} we show the three equivalent width ratios defined by \citet{sota11,sota14} to refine spectral classification among O8-O9.7 stars. The values of the average ratios are gathered in Table \ref{tab_ave_st89}. We see that for the three ratios there is a clear trend with spectral type. Both \ion{He}{i}~4388/\ion{He}{ii}~4542 and \ion{He}{i}~4144/\ion{He}{ii}~4200 are larger than 1.0 for O9.5 and O9.7 stars, and lower than 1.0 for O8 stars. For O8.5, O9 and O9.2, the ratios are close to 1.0 and the range of observed values overlap. \ion{He}{i}~4144/\ion{He}{ii}~4200 shows a stronger variation than \ion{He}{i}~4388/\ion{He}{ii}~4542 across these spectral types and may be preferred to separate between them. \citet{sota11} define spectral type O9 as the one where \ion{He}{i}~4144/\ion{He}{ii}~4200 and \ion{He}{i}~4388/\ion{He}{ii}~4542 are about 1.0: our results are consistent with this definition. There is a large overlap in the equivalent width ratios at spectral types O9.5 and O9.7. The main difference is that much larger values (up to between three and five) are reached at O9.7, whereas the ratios are limited to $\lesssim$2 at O9.5.
The ratio \ion{Si}{iii}~4552/\ion{He}{ii}~4542 follows a monotonic variation in the spectral range covered, being almost 0.0 at O8 and reaching $\sim$ 1.0 at O9.7 which corresponds to the definition of the O9.7 spectral type \citep{sota11}.

These findings are consistent with the qualitative classification proposed by \citet{sota14} -- see their Table 3. The only difference is the separation between O9 and O9.2: we find that the two sub-classes can be distinguished only from \ion{He}{i}~4144/\ion{He}{ii}~4200. For the other two ratios, there is no significant difference. But we also stress that the number of O9.2 stars in our sample remains small (six, versus thirteen O9 stars) and thus this result may be revised in future studies.
We note that our ratios are expressed in terms of equivalent widths, while \citet{sota11,sota14} define ratios of peak intensities. A test of the effect of these different definitions on the ratio \ion{He}{i}~4144/\ion{He}{ii}~4200 reveals that the same trends are observed if peak intensities are used instead of equivalent widths.

\begin{table}
\begin{center}
\caption{Average and dispersion of ratios shown in Fig.\ \ref{ew_st89_obs} and defining spectral types between O8 and O9.7.}
\label{tab_ave_st89}
\begin{tabular}{lccc}
\hline
\vspace{0.1cm} 
  Spectral type  & $\frac{EW(HeI 4144)}{EW(HeII 4200)}$ & $\frac{EW(HeI 4388)}{EW(HeII 4542)}$ & $\frac{EW(SiIII 4552)}{EW(HeII 4542)}$\\
\hline
O8    & 0.38$\pm$0.14 & 0.55$\pm$0.25 & 0.07$\pm$0.04  \\
O8.5  & 0.67$\pm$0.15 & 0.96$\pm$0.17 & 0.13$\pm$0.04  \\
O9    & 0.85$\pm$0.28 & 1.08$\pm$0.20 & 0.28$\pm$0.09  \\
O9.2  & 1.11$\pm$0.27 & 1.14$\pm$0.13 & 0.31$\pm$0.09  \\
O9.5  & 1.37$\pm$0.32 & 1.53$\pm$0.40 & 0.39$\pm$0.17  \\
O9.7  & 1.98$\pm$1.00 & 2.45$\pm$0.99 & 0.96$\pm$0.57  \\
\hline
\end{tabular}
\end{center}
\end{table}

\subsection{Luminosity class}
\label{s_lc}

\subsubsection{O3-O7.5 stars}
\label{s_lc_early}

\begin{figure}[]
\centering
\includegraphics[width=9cm]{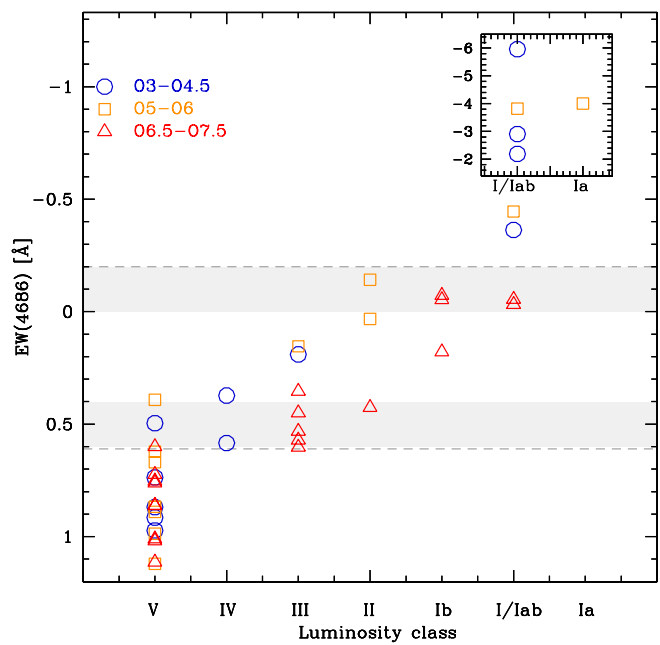}
\caption{Equivalent width of \ion{He}{ii}~4686 as a function of luminosity class for stars with spectral type between O3 and O7.5. Gray dashed lines show the proposed limit between luminosity classes. Light gray shaded areas show the overlapping regions between various luminosity classes.}
\label{ew4686_lc_obs}
\end{figure}

The main luminosity class criterion for stars with spectral type earlier than O7.5 is the strength of \ion{He}{ii}~4686. It is usually accompanied by considering the strength of the \ion{N}{iii}~4634-40-42 emission complex. The morphology of both the \ion{He}{ii} and \ion{N}{iii} lines also defines the ``f'' phenomenon: ((f)) stars have weak nitrogen emission and strong helium absorption; (f) stars have medium nitrogen emission and neutral or weak helium absorption; f stars have strong nitrogen and helium emission.

In Fig.\ \ref{ew4686_lc_obs} we show the equivalent width of \ion{He}{ii}~4686 for stars of different luminosity classes.
All stars with an equivalent width larger than 0.60 \AA\ are dwarfs. Equivalently, all stars with EW(\ion{He}{ii}~4686) smaller than -0.20 \AA\ (i.e., emission stronger than 0.20 \AA) are supergiants of class I/Iab or Ia. We thus propose these limits to uniquely (but not exclusively) define dwarfs and supergiants. Gray shaded areas in Fig.\ \ref{ew4686_lc_obs} indicate transition regions between dwarfs and giants, and between giants and supergiants. More precisely, there is no dwarf with EW(\ion{He}{ii}~4686) smaller than 0.40 \AA. In the range 0.40--0.60 \AA, stars can be dwarfs or sub-giants, giants, or bright giants. Similarly, all but one\footnote{HD~156154, a O7.5Ib star, has  EW(\ion{He}{ii}~4686)=0.18 \AA} of the bright giants and supergiants have negative EW(\ion{He}{ii}~4686), meaning that they have \ion{He}{ii}~4686 mainly in emission. Hence stars with EW(\ion{He}{ii}~4686) between 0.00 and -0.20~\AA\ can be bright giants or supergiants. For stars in these two transition regions (0.40--0.60~\AA\ and -0.20--0.00~\AA), we suggest a direct comparison to standard spectra for a final luminosity class determination.

From Fig.\ \ref{ew4686_lc_obs} we note that for giants and supergiants EW(\ion{He}{ii}~4686) is on average smaller (i.e., more emission) among O3-O6 stars compared to O6.5-O7.5 stars. For instance, there is no O6.5-O7.5 giant (luminosity class IV, III and II) with EW smaller than $\sim$0.30 \AA, while almost all the O3-O6 giants have smaller EWs. Similarly there is no O6.5-O7.5 supergiant with EW smaller than -0.20 \AA\ while all O3-O6 supergiants have more emission.

\begin{figure*}[]
\centering
\includegraphics[width=0.49\textwidth]{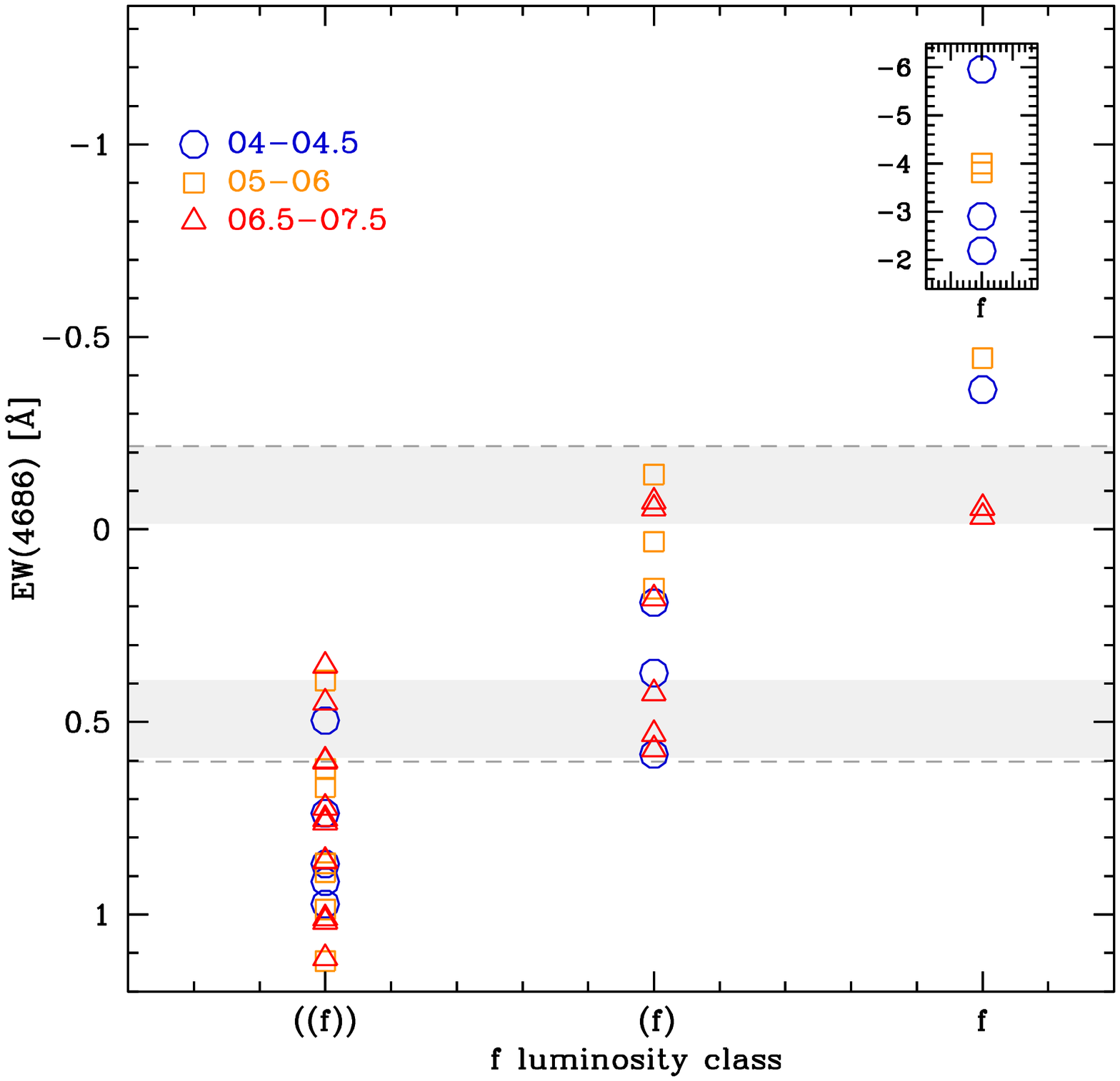}
\includegraphics[width=0.49\textwidth]{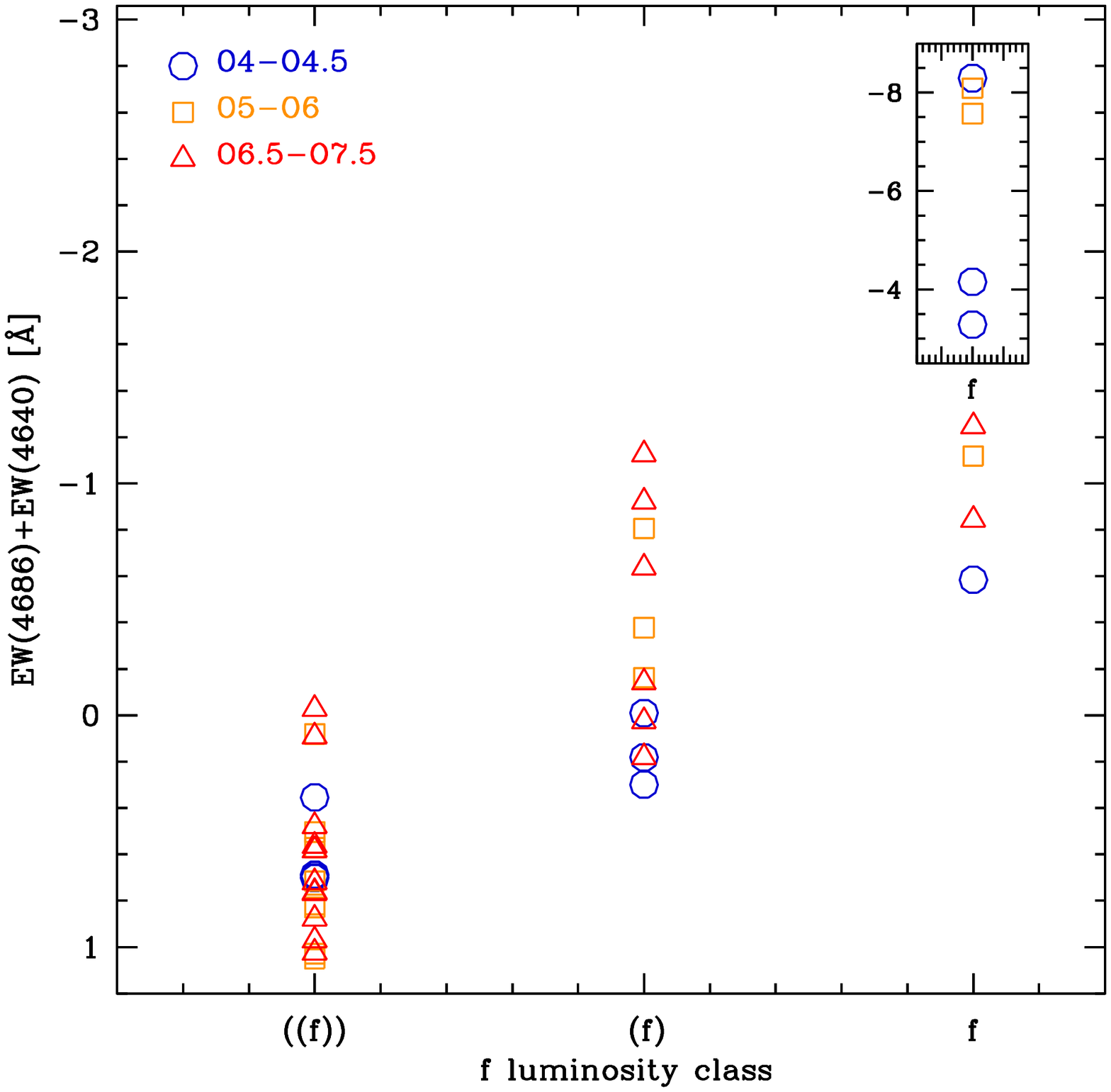}
\caption{Equivalent width of \ion{He}{ii}~4686 (left panel) and \ion{He}{ii}~4686 + \ion{N}{iii}~4634-40-42 (right panel) as a function of ``f'' criterion for stars with spectral type between O4 and O7.5. In the left panel gray dashed lines and shaded areas are the same as in Fig.\ \ref{ew4686_lc_obs}.}
\label{ew4686_f_obs}
\end{figure*}

Interestingly, the equivalent width ranges just discussed define remarkably well the three ((f)), (f) and f categories if one considers only EW(\ion{He}{ii}~4686), as illustrated in the left panel of Fig.\ \ref{ew4686_f_obs}. All stars with an equivalent width larger than 0.60 \AA\ (respectively smaller than -0.20 \AA) have the ((f)) (respectively the f) classification. Stars classified has (f) objects have EW exactly in between these two values. The range 0.00--0.40 \AA\ is populated uniquely by (f) stars. For EW(\ion{He}{ii}~4686) between 0.40 and 0.60~\AA\ there are both ((f)) and (f) stars while for EW(\ion{He}{ii}~4686) between 0.00 and -0.20 \AA, there are (f) and f stars (gray shaded areas in Fig.\ \ref{ew4686_f_obs}.) 
When taking the full definition of the ``f'' phenomenon (right panel of Fig.\ \ref{ew4686_f_obs}) the three groups of stars remain well separated. From these two figures one concludes that the ``f'' phenomenon can be entirely defined by just the equivalent width of \ion{He}{ii}~4686, and that the ranges in EW(\ion{He}{ii}~4686) defining the luminosity classes V, IV-III-II (grouped) and I can also be used to define the ((f)), (f) and f categories. We thus advise the following quantitative classification scheme:

\begin{itemize}

\item For stars with EW(\ion{He}{ii}~4686) $>$ 0.60, a luminosity class V together with a ((f)) qualifier can be assigned.

\item For stars with EW(\ion{He}{ii}~4686) $<$ -0.20, a luminosity class I (Iab or Ia) and a f qualifier can be assigned.

\item For stars with -0.20 $<$ EW(\ion{He}{ii}~4686) $<$ 0.60, various luminosity classes and ``f'' qualifier can be assigned with the restriction that EWs between 0.00 and 0.40 \AA\ correspond uniquely to the (f) qualifier.
  
\end{itemize}

\noindent Clearly for equivalent widths close to the above limits, special care must be taken when assigning a luminosity class and comparison to standard stars spectra is advised. Our proposed scheme is generally consistent with Table~2 of \citet{sota14} which relates the f qualifier to luminosity classes for various spectral types.

\subsubsection{O8-O8.5 stars}
\label{s_lc_O8}

\begin{figure}[ht]
\centering
\includegraphics[width=0.49\textwidth]{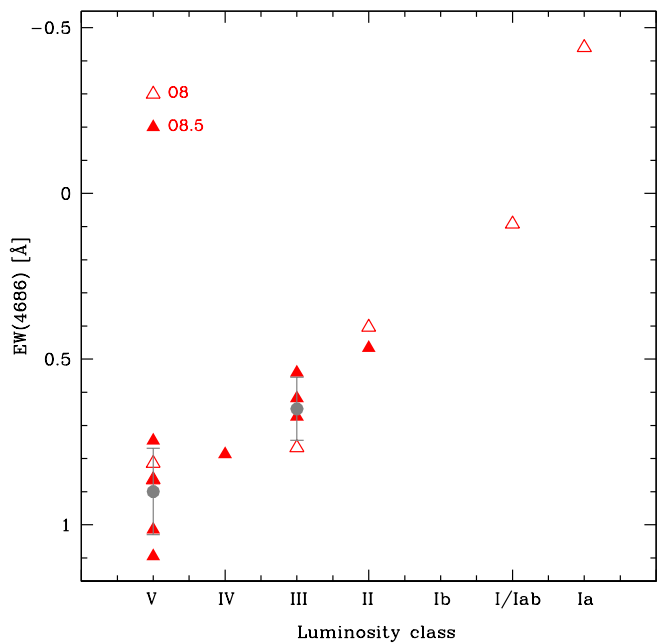}
\caption{Equivalent width of \ion{He}{ii}~4686 as a function of luminosity class for O8-O8.5 stars. Gray circles and error bars indicate the average and standard deviation for luminosity classes V and III.}
\label{lc_O8}
\end{figure}

At spectral types O8 and O8.5 the strength of \ion{He}{ii}~4686 remains the main luminosity class criterion. However, the mass loss rates are lower than at earlier spectral types so that the strength of \ion{He}{ii}~4686 is weaker for a given luminosity class, as was already the case between O3-O6 and O6.5-O7.5 (see above). In addition, the ``f'' qualifier is seldom used for O8-O8.5 stars. Consequently, a dedicated luminosity classification scheme is defined for these spectral types.
In Fig.\ \ref{lc_O8} we show the equivalent width of \ion{He}{ii}~4686 as a function of luminosity class for the fifteen O8-O8.5 stars of our sample. There is a clear trend of less absorption/more emission when moving from dwarfs to supergiants. We note the small number of stars in the supergiant classes. This is mainly due to the very small number of O8-O8.5 supergiants known: in the catalog of \citet{ma16} there are only seven (six) O8 (O8.5) supergiants (of all subclasses Ia, Iab, and Ib). The O8-O8.5 supergiants are thus a rare class of objects. For these reasons we provide the average value of EW(\ion{He}{ii}~4686) only for dwarfs and giants: 0.90$\pm$0.13 for the former, 0.65$\pm$0.10 for the latter.

\subsubsection{O9-O9.7 stars}
\label{s_lc_late}

\begin{figure*}[h]
\centering
\includegraphics[width=0.49\textwidth]{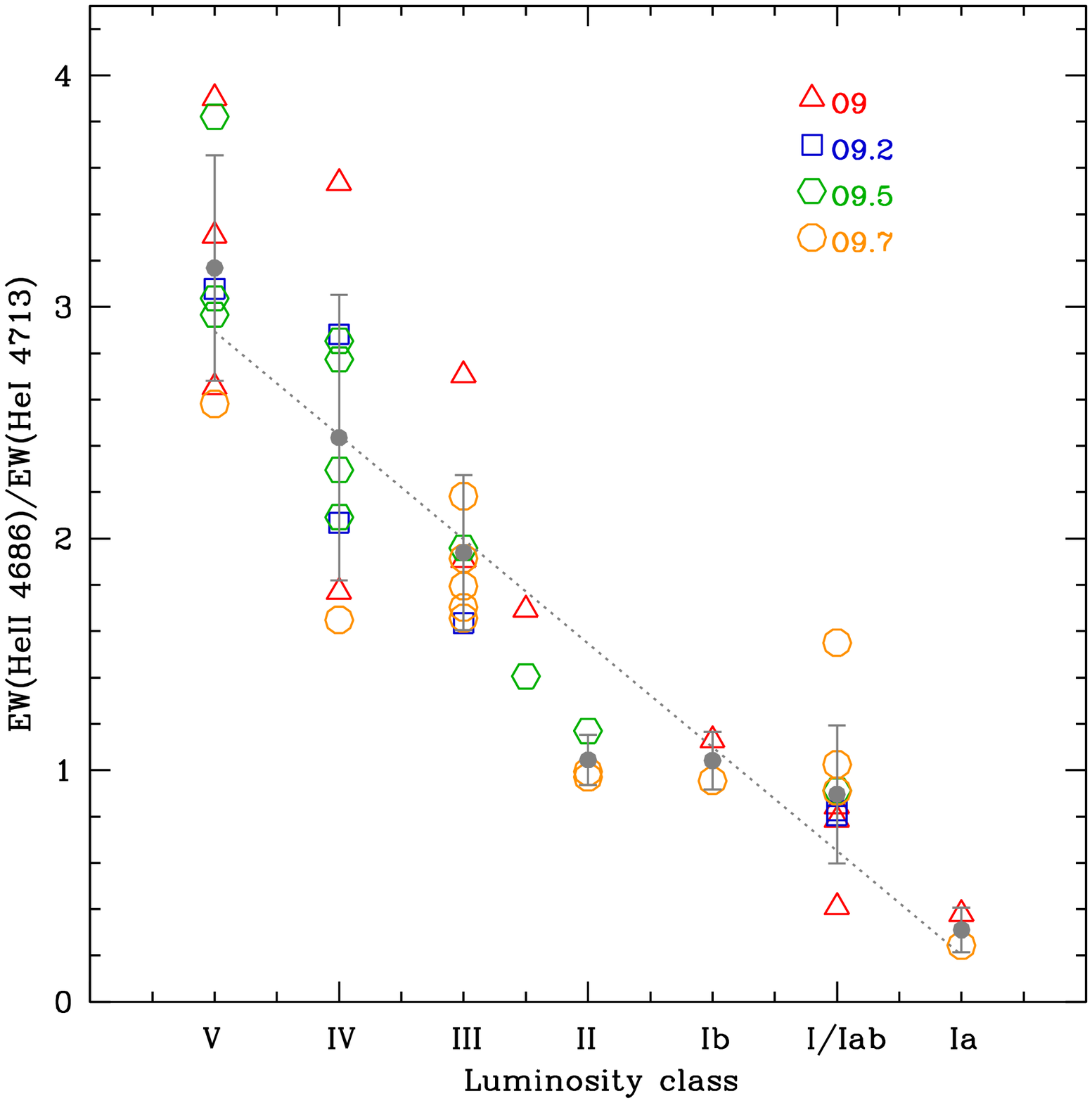}
\includegraphics[width=0.49\textwidth]{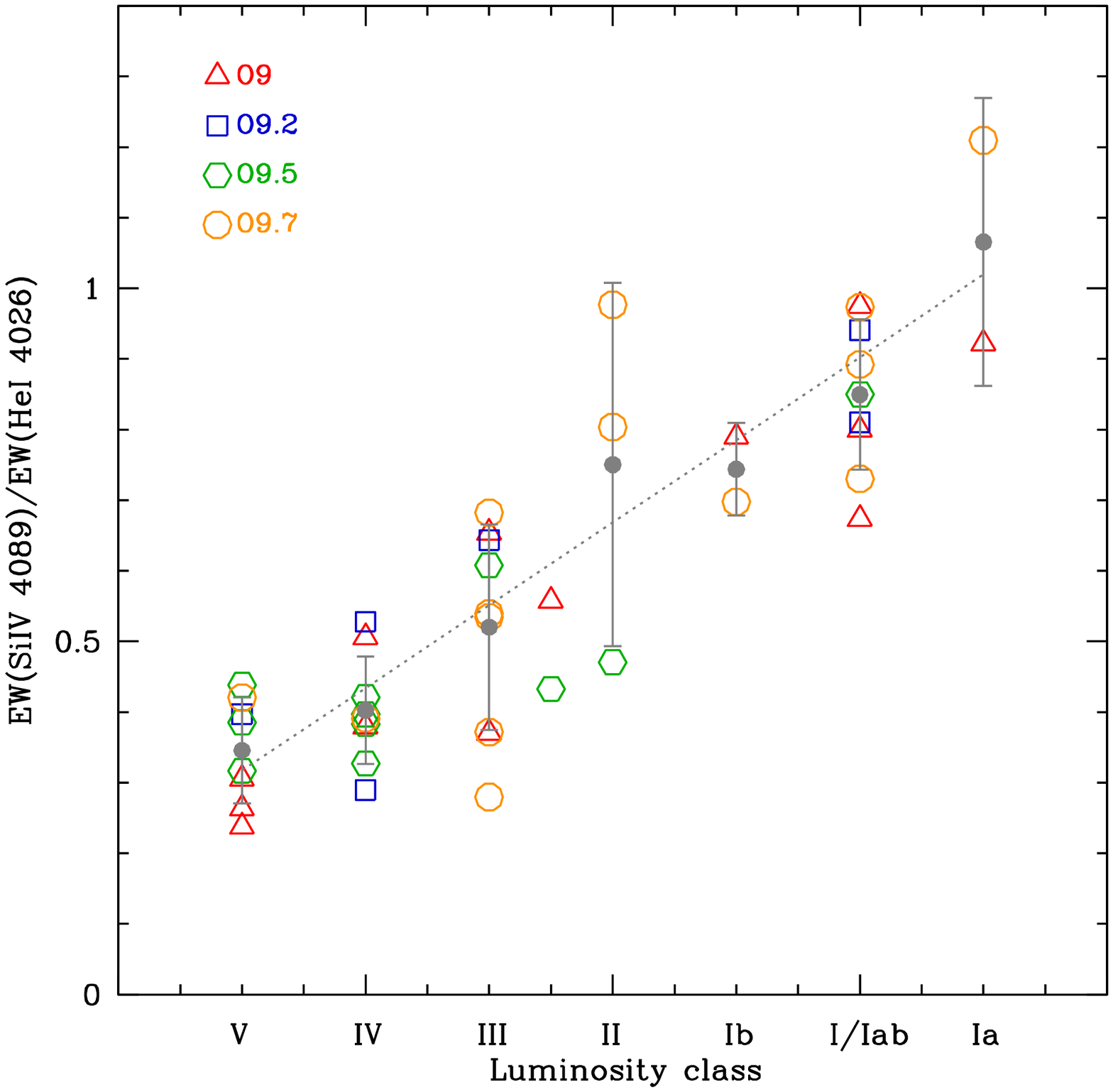}
\caption{Luminosity class criterion for O9-O9.7 stars. \textit{Left:} Ratio of the equivalent widths of \ion{He}{ii}~4686 to \ion{He}{i}~4713 as a function of luminosity class. \textit{Left:} Ratio of the equivalent widths of \ion{Si}{iv}~4089 to \ion{He}{i}~4026 as a function of luminosity class. Gray circles and error bars indicate the average and standard deviation for each luminosity class. Dotted lines are the least square fits to the average ratios.}
\label{lc_O9}
\end{figure*}

Luminosity classes in the O9-O9.7 spectral type range are best defined using the following two ratios: $\frac{EW(\ion{He}{ii}~4686)}{EW(\ion{He}{i}~4713)}$ and $\frac{EW(\ion{Si}{iv}~4089)}{EW(\ion{He}{i}~4026)}$ \citep{sota11,sota14}. These ratios are shown in Fig.\ \ref{lc_O9}. For both ratios, there is a clear variation among the entire luminosity class range. The former varies by a factor of about ten between dwarfs and type Ia supergiants. For the latter, the variation is of a factor $\sim$3. Although the ranges covered by a given ratio usually overlap between neighboring luminosity classes, the average values show monotonic variations (with the exception of the II class, due to the small number of objects). We note the absence of differences between spectral types: O9 stars cover almost the same range of ratios as O9.5 stars. However, we stress that this is likely due to small number statistics. Future studies based on hundreds of spectra may reveal variations unseen in our sample. 

\citet{sota11} indicate that a ratio of about one in the ratio of the intensities of \ion{He}{ii}~4686 to \ion{He}{i}~4713 is reached for bright giants (luminosity class II) with small rotational broadening. In our case, where we use equivalent widths, the ratio of 1.0 corresponds more to Ib stars. We stress that we prefer the ratio of equivalent width since in our sample the ratio based on intensities is basically the same for luminosity classes V, IV and III. It is thus less discriminant. 

We advise to use Fig.\ \ref{lc_O9} as a first guess for luminosity class assignment. After computing the observed ratios, they can be plotted on top of Fig.\ \ref{lc_O9} to assign a luminosity class. Table \ref{tab_ave_lc9} also gathers the average values (and dispersion) of the ratios shown in Fig.\ \ref{lc_O9}. Final luminosity class assignment or verification should rely on comparison to standard stars spectra. 

Finally we note that \citet{ca71} used the ratio $\frac{EW(\ion{Si}{iv}~4089)}{EW(\ion{He}{i}~4144)}$ for luminosity class determination. We found that this ratio is indeed correlated with luminosity class, but in our measurements the dispersion is much larger than for $\frac{EW(\ion{Si}{iv}~4089)}{EW(\ion{He}{i}~4026)}$. We thus prefer to rely on the latter ratio, as also advocated by \citet{sota11}. \citet{sota14} also note that the ratios \ion{He}{ii}~4686 to \ion{He}{i}~4713 and  \ion{Si}{iv}~4089 to \ion{He}{i}~4026 can lead to discrepant results, especially in the case of unresolved spectroscopic binaries. In Fig.~\ref{lc_O9} HD~225146, a O9.7Iab star according to \citet{sota11}, has a high $\frac{EW(\ion{He}{ii}~4686)}{EW(\ion{He}{i}~4713)}$ ratio that would better correspond to a luminosity class II (or marginally III) according to our results. Its $\frac{EW(\ion{Si}{iv}~4089)}{EW(\ion{He}{i}~4026)}$ ratio is also more consistent with a II or Ib star. This shows that, as stated by \citet{sota14}, different line ratios can lead to different luminosity classes. In addition, our quantitative criteria, if used to reassign luminosity classes, can lead to revision compared to the qualitative classification processes. In the case of HD~225146 one would probably assign a luminosity class II using Fig.\ \ref{lc_O9} instead of Iab from comparison to standard stars. This stresses that spectral classification can have low precision due to the diversity of criteria and morphology among spectral features.

\begin{figure}[h]
\centering
\includegraphics[width=0.49\textwidth]{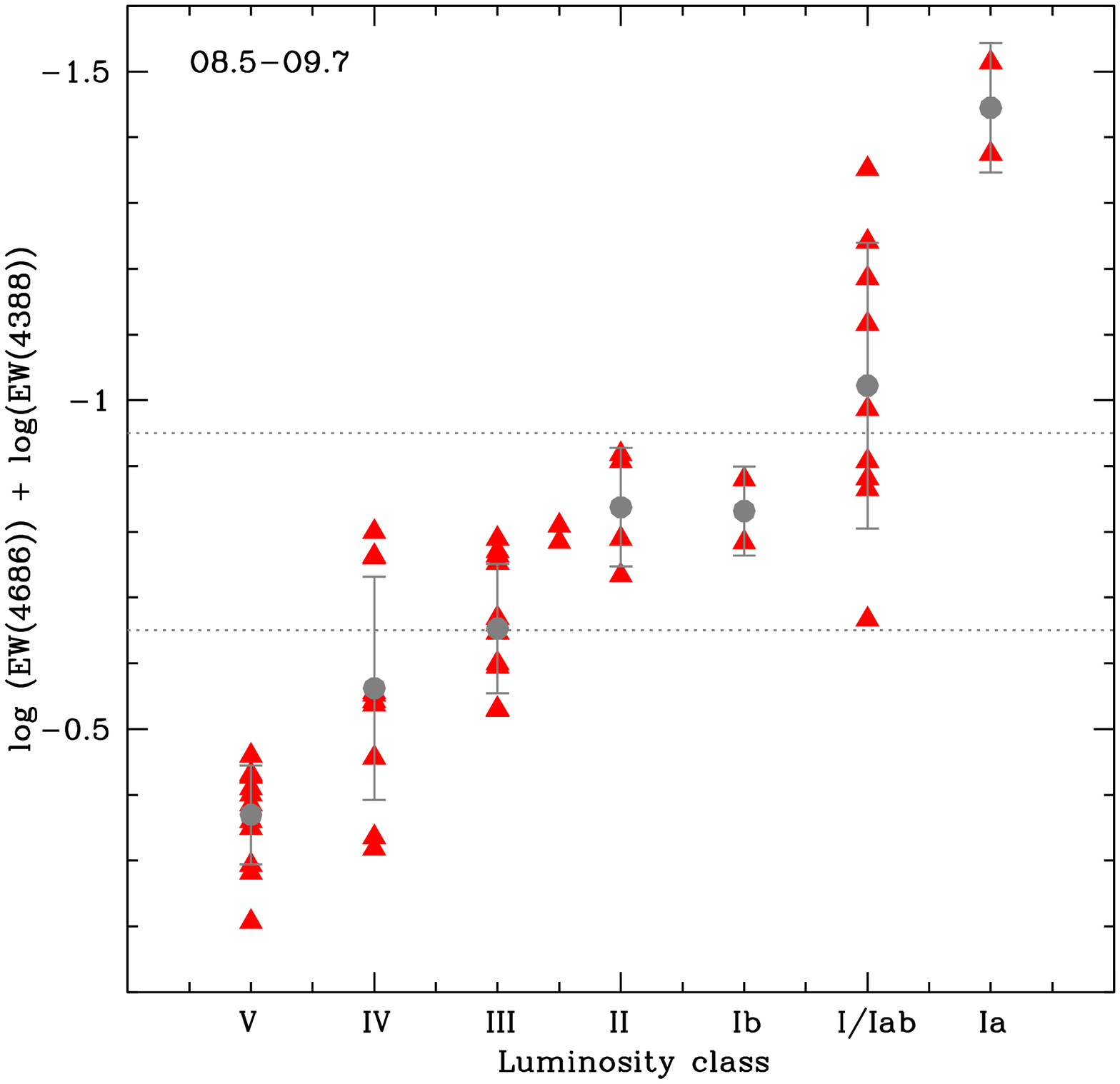}
\caption{Logarithm of the equivalent width of \ion{He}{ii}~4686 plus logarithm of the equivalent width of \ion{He}{i}~4388 as a function of luminosity class for O8.5-O9.7 stars. Gray circles and error bars indicate the average and standard deviation. The dotted lines indicates the limits defined by \citet{mathys89} for luminosity classes V, III and I. We note that we use \AA\ for the unity of equivalent width measurements, while \citet{mathys89} used m\AA.}
\label{lc_mathys}
\end{figure}

\begin{table}
\begin{center}
\caption{Average and dispersion of ratios shown in Fig.\ \ref{lc_O9} and defining luminosity classes for spectral types O9-O9.7.}
\label{tab_ave_lc9}
\begin{tabular}{lccc}
\hline
\vspace{0.1cm} 
Luminosity Class & $\frac{EW(HeII 4686)}{EW(HeI 4713)}$ & $\frac{EW(SiIV 4089)}{EW(HeI 4026)}$ \\
\hline
V     & 3.17$\pm$0.49  & 0.35$\pm$ 0.07  \\
IV    & 2.44$\pm$0.62  & 0.40$\pm$ 0.08  \\
III   & 1.94$\pm$0.33  & 0.52$\pm$ 0.15  \\
II    & 1.04$\pm$0.11  & 0.75$\pm$ 0.26  \\
Ib    & 1.04$\pm$0.12  & 0.74$\pm$ 0.07  \\
I/Iab & 0.90$\pm$0.30  & 0.85$\pm$ 0.11  \\
Ia    & 0.31$\pm$0.10  & 1.07$\pm$ 0.20  \\
\hline
\end{tabular}
\end{center}
\end{table}

\smallskip

For completeness we show in Fig.\ \ref{lc_mathys} the luminosity class criterion for O8.5-O9.7 stars as defined by \citet{mathys89}. There is a clear trend with luminosity class. In particular, luminosity classes V, III, and Iab-Ia are well separated. For other luminosity classes, overlap with neighboring classes is significant. For instance, luminosity class II stars have $\log(\ion{He}{ii}~4686)+\log(\ion{He}{i}~4388)$ that can be those of giants or Ib stars. 
The limits defined by \citet{mathys89} are only partly consistent with our measurements. Dwarfs fall in the Mathys range, but giants (LC III) and supergiants (except Ia) overlap between the V--III and III-I ranges of Mathys. We may tentatively attribute this discrepancy to the limited spectral resolution of the Mathys spectra (R$\sim$5200).

\section{Conclusion}
\label{s_conc}

We have collected high-resolution optical spectra of 105 Galactic O-type stars from public archives (\emph{ELODIE}, \emph{SOPHIE}, CFHT Science, ESO and PolarBase archives). We have measured equivalent widths of key classification lines. We have calculated the average and dispersion of classification criteria for given spectral types and luminosity classes. We have defined ranges of equivalent widths of \ion{He}{ii}~4686 for various luminosity classes among O3-O8.5 stars and for the ``f'' phenomenon. We have presented quantative spectral type and luminosity class criteria for O8-O9.7 stars. The tabulated values should be used for spectral classification prior to a final comparison to standard stars spectra of the assigned spectral type and luminosity class. Our results should be improved and refined in the future when hundreds of high-resolution spectra of O stars currently collected by spectroscopic surveys \citep{barba10,ss15} are available.

\begin{acknowledgements}
We acknowledge a timely and construtive report from the referee, J. Ma\'iz-Apell\'aniz. Many thanks to R. Barb\'a for allowing the use of \emph{FEROS} archive spectra prior to publication and following the referee's suggestion. This research used the \emph{ELODIE} and \emph{SOPHIE} archives developed and maintained by OHP/INSU-CNRS/OSU Pytheas; the PolarBase archive developed and maintained by OMP/INSU-CNRS/Toulouse University; the facilities of the Canadian Astronomy Data Centre operated by the National Research Council of Canada with the support of the Canadian Space Agency. This research has made use of the services of the ESO Science Archive Facility.
\end{acknowledgements}

\bibliographystyle{aa}
\bibliography{classif}

\begin{appendix}
  
\section{Observational material}
\label{ap_obs}

\begin{table}
\begin{center}
\caption{Sample stars, spectral type (ST), luminosity class (LC) and spectrograph used for data acquisition.}
\label{tab_obs}
\begin{tabular}{lllll}
\hline
Star         & ST       & LC     & Additional & Spectrograph\\
             &          &        & qualifier\\
\hline
HD 64568      &  3      &  V     &     ((f*))z    &  FEROS \\
HD 319718     &  3.5    &  I     &      f*        &  FEROS \\
HD 93205      &  3.5    &  V     &      ((f))     &  FEROS \\
HD 15570      &  4      &  I     &      f         &  SOPHIE \\  
HD 16691      &  4      &  I     &      f         &  ELODIE\\  
HD 66811      &  4      &  I     &      (n)fp     &  FEROS\\
HD 93250      &  4      &  IV    &      (fc)      &  FEROS\\
HD 164794     &  4      &  V     &      ((f))     &  FEROS\\
HD 96715      &  4      &  V     &      ((f))z    &  FEROS\\
HD 46223      &  4      &  V     &      ((f))     &  SOPHIE\\  
HD 15558      &  4.5    &  III   &      (f)       &  ELODIE\\  
HD 192281     &  4.5    &  IV    &      (n)(f)    &  SOPHIE\\
ALS 2063      &  5      &  I     &      fp        &  FEROS\\
CPD -47 2963  &  5      &  I     &      fc        &  FEROS\\
HD 93843      &  5      &  III   &      (fc)      &  FEROS\\
HD 46150      &  5      &  V     &      ((f))     &  SOPHIE\\
HD 319699     &  5      &  V     &      ((fc))    &  FEROS\\
BD -16 4826   &  5.5    &  V     &      ((f))z    &  FEROS\\    
CPD -59 2673  &  5.5    &  V     &      (n)((f))z &  FEROS\\ 
HD 93204      &  5.5    &  V     &      ((f))     &  FEROS\\
HD 153919     &  6      &  Ia    &      fcp       &  ESPaDOnS\\  
HDE 229196    &  6      &  II    &      (f)       &  SOPHIE\\
HD 152233     &  6      &  II    &      (f))      &  FEROS\\
CPD -59 2600  &  6      &  V     &      ((f))     &  FEROS\\
HD 303311     &  6      &  V     &      ((f))z    &  FEROS\\ 
HD 42088      &  6      &  V     &      ((f))z    &  SOPHIE\\  
HD 190864     &  6.5    &  III   &      (f)       &  SOPHIE\\ 
HD 17505      &  6.5    &  III   &      n(f)      &  ELODIE\\ 
HD 12993      &  6.5    &  V     &      ((f))Nstr &  SOPHIE\\
HD 228841     &  6.5    &  V     &      n((f))    &  SOPHIE\\ 
HD 199579     &  6.5    &  V     &      ((f))z    &  ESPaDOnS\\ 
HD 193514     &  7      &  Ib    &      (f)       &  SOPHIE\\
HDE 242926    &  7      &  V     &      z         &  SOPHIE\\ 
HD 47839      &  7      &  V     &      ((f))z    &  ESPaDOnS\\ 
HD 188001     &  7.5    &  Iab   &      f         &  ESPaDOnS\\
HD 192639     &  7.5    &  Iab   &      f         &  SOPHIE\\
HD 17603      &  7.5    &  Ib    &      (f)       &  SOPHIE\\ 
HD 156154     &  7.5    &  Ib    &      (f)       &  ESPaDOnS\\ 
HD 34656      &  7.5    &  II    &      (f)       &  SOPHIE\\ 
HD 186980     &  7.5    &  III   &      ((f))     &  ESPaDOnS\\
HD 203064     &  7.5    &  III   &      n((f))    &  ESPaDOnS\\
HD 24912      &  7.5    &  III   &      (n)((f))  &  SOPHIE\\
HD 35619      &  7.5    &  V     &      ((f))     &  ESPaDOnS\\ 
HD 36879      &  7.5    &  V     &      (n)((f))  &  ESPaDOnS\\
HD 155806     &  7.5    &  V     &      ((f))     &  ESPaDOnS\\ 
HD 164492     &  7.5    &  V     &      z         &  ESPaDOnS\\
HD 151804     &  8      &  Ia    &      f         &  HARPSpol\\
HD 225160     &  8      &  Iab   &      f         &  SOPHIE\\ 
HD 162978     &  8      &  II    &      ((f))     &  ESPaDOnS\\              
HD 36861      &  8      &  III   &      ((f))     &  ESPaDOnS\\
HD 191978     &  8      &  V     &                &  SOPHIE\\ 
HD 66788      &  8      &  V     &                &  ESPaDOnS\\
\hline                                                                      
\end{tabular}                                                               
\end{center}                                                                
\end{table}

\setcounter{table}{0}

\begin{table}
\begin{center}
\caption{Continued}
\label{tab_obs2}
\begin{tabular}{lllll}
\hline
Star         & ST       & LC     & Additional & Spectrograph\\
             &          &        & qualifier\\  
\hline
HD 207198     &  8.5    &  II    &      ((f))     &  SOPHIE\\ 
HD 218195     &  8.5    &  III   &                &  SOPHIE\\
HD 153426     &  8.5    &  III   &                &  ESPaDOnS\\
HD 13268      &  8.5    &  III   &                &  ELODIE\\
HD 46966      &  8.5    &  IV    &                &  SOPHIE\\
HD 46149      &  8.5    &  V     &                &  SOPHIE\\
HD 14633      &  8.5    &  V     &                &  SOPHIE\\
HD 216532     &  8.5    &  V     &      (n)       &  ELODIE\\ 
HD 48279      &  8.5    &  V     &                &  ELODIE\\
HD 30614      &  9      &  Ia    &                &  SOPHIE\\
HD 152249     &  9      &  Iab   &                &  ESPaDOnS\\
HD 202124     &  9      &  Iab   &                &  SOPHIE\\
HD 210809     &  9      &  Iab   &                &  ESPaDOnS\\
HD 209975     &  9      &  Ib    &                &  SOPHIE\\
HD 16429      &  9      &  III-II&       (n)      &  ELODIE\\
HD 193443     &  9      &  III   &                &  ESPaDOnS\\
HD 24431      &  9      &  III   &                &  SOPHIE\\
HD 193322     &  9      &  IV    &       (n)      &  ESPaDOnS\\ 
HD 209481     &  9      &  IV    &       (n)      &  ESPaDOnS\\ 
HD 214680     &  9      &  V     &                &  SOPHIE\\
HD 207898     &  9      &  V     &                &  ELODIE\\
HD 258691     &  9      &  V     &                &  ESPaDOnS\\
HD 154368     &  9.2    &  Iab   &                &  ESPaDOnS\\
HD 218915     &  9.2    &  Iab   &                &  ESPaDOnS\\
HD 152247     &  9.2    &  III   &                &  ESPaDOnS\\
HD 149757     &  9.2    &  IV    &       nn       &  NARVAL\\ 
HD 201345     &  9.2    &  IV    &                &  ESPaDOnS\\
HD 46202      &  9.2    &  V     &                &  ESPaDOnS\\
HD 188209     &  9.5    &  Iab   &                &  SOPHIE\\
HD 36486      &  9.5    &  II    &                &  ELODIE\\
HD 15137      &  9.5    &  III-II&       n        &  ELODIE\\ 
HD 167263     &  9.5    &  III   &                &  ESPaDOnS\\
HD 192001     &  9.5    &  IV    &                &  SOPHIE\\
HD 36483      &  9.5    &  IV    &                &  ELODIE\\
HD 155889     &  9.5    &  IV    &                &  HARPSpol\\
HD 206183     &  9.5    &  V-IV  &                &  ESPaDOnS\\
BD+60 499     &  9.5    &  V     &                &  ESPaDOnS\\
HD 34078      &  9.5    &  V     &                &  SOPHIE\\
HD 38666      &  9.5    &  V     &                &  SOPHIE\\
HD 195592     &  9.7    &  Ia    &                &  SOPHIE\\
HD 167264     &  9.7    &  Iab   &                &  NARVAL\\
HD 225146     &  9.7    &  Iab   &                &  SOPHIE\\
HD 149038     &  9.7    &  Iab   &                &  ESPaDOnS\\
HD 47432      &  9.7    &  Ib    &                &  SOPHIE\\
HD 328856     &  9.7    &  II    &                &  ESPaDOnS\\
HD 13745      &  9.7    &  II    &       (n)      &  NARVAL\\ 
HD 55879      &  9.7    &  III   &                &  ESPaDOnS\\
HD 46106      &  9.7    &  III   &       (n)      &  ESPaDOnS\\ 
HD 189957     &  9.7    &  III   &                &  SOPHIE\\
HD 37468      &  9.7    &  III   &                &  ESPaDOnS\\
HD 154643     &  9.7    &  III   &                &  ESPaDOnS\\
HD 207538     &  9.7    &  IV    &                &  SOPHIE\\
HD 36512      &  9.7    &  V     &                &  NARVAL\\
\hline                                                                      
\end{tabular}                                                               
\end{center}                                                                
\end{table}

\end{appendix}                                                              
                                                                            
\end{document}